\documentclass[groupedaddress,prl,twocolumn,showpacs,amsmath,amssymb, amsfonts,10pt,superscriptaddress,noeprint]{revtex4-2}

\usepackage[T1]{fontenc}
\usepackage{graphicx}
\usepackage{enumitem}
\usepackage{amsmath}
\usepackage{amsfonts}
\usepackage{amssymb}
\usepackage{mwe}
\usepackage{color}
\usepackage[dvipsnames]{xcolor}
\usepackage{bm}
\usepackage{soul}
\usepackage[normalem]{ulem}
\usepackage[colorlinks,bookmarks=false,urlcolor=blue,citecolor=blue,linkcolor=blue]{hyperref}
\usepackage[capitalise]{cleveref}
\usepackage{comment}
\usepackage[normalem]{ulem}

\definecolor{darkgreen}{rgb}{0,0.5,0}
\definecolor{orange}{rgb}{1,0.5,.3}
\definecolor{darkred}{rgb}{.7,0,0}
\definecolor{purple}{rgb}{0.6,0,0.5}
\definecolor{darkpetrol}{RGB}{0,73,76}

\usepackage[normalem]{ulem}

\renewcommand {\phi}{{\varphi}}

\begin{document}
\title{
{Enhancing superconductivity using thermal bosons}
}

\author{Ekaterina Vlasiuk}
\affiliation{Institute for Theoretical Physics, Heidelberg University, 69120 Heidelberg, Germany}

\author{Manfred Salmhofer}
\affiliation{Institute for Theoretical Physics, Heidelberg University, 69120 Heidelberg, Germany}

\author{Eugene Demler}
\affiliation{Institute for Theoretical Physics, ETH Z\"urich, 8093 Z\"urich, Switzerland}

\author{Richard Schmidt}
\affiliation{Institute for Theoretical Physics, Heidelberg University, 69120 Heidelberg, Germany}

\begin{abstract}

We investigate how the strong coupling of a superconductor to thermal bosons can enhance its  superconducting critical temperature. To tackle this problem, we use a renormalization group approach that allows us to describe the competition between density fluctuations and the build-up of boson-induced attraction between fermions. Capturing the mutual influence of bosonic and fermionic sectors, the self-consistent renormalization group framework predicts a robust increase of the critical temperature across a wide range of interactions. We find a nontrivial dependence of the critical temperature on the boson mass and we establish a phase diagram for enhanced superconductivity driven by bosons being either in the condensed or thermal state. We outline possible experimental realizations in cold atomic systems and discuss  implementations using electron-exciton mixtures in van der Waals material heterostructures.

\end{abstract}
\maketitle

What are possible ways to increase the critical temperature $T_c$ of a  superconductor? Are there bounds on $T_c$ and what defines them? Can bounds be changed by coupling the superconductor (SC) to a separate many-body system that may or may not be in a quantum degenerate state? These are among the questions that have recently come into focus driven by theory~\cite{Hofmann2022, Esterlis2018, Hazra2019} and experimental progress alike~\cite{Regal2004, Ries2015, Oh2021, Fausti2011}. In this letter, we consider these questions by studying how the critical temperature of a generic superconductor reacts to the coupling to a thermal bosonic medium. In particular we explore how a  significant enhancement of $T_c$ can be achieved when entering the strong-coupling regime, where  electrons may form bound states with the bosons, such as realized in two-dimensional (2D) materials~\cite{Sidler2017} or synthetic quantum matter~\cite{Duda2023,Cai2026}. 

We focus on two aspects, namely whether  a purely thermal medium can enhance $T_c$, and whether it is possible to transition from one mechanism producing a bound on $T_c$ to another.  To gain intuition, consider a  phonon-mediated superconductor: here the Debye energy acts as the natural scale that limits $T_c$~\cite{Esterlis2018}. But what happens if one couples this superconductor to an additional species of particles? The question then arises whether  the superconductor's internal bound on $T_c$  is still the  relevant one, or if there is a transition to another bound associated with the scales intrinsic to the newly-introduced medium.

While  qualitative proposals for exciton-mediated superconductors \cite{ Ginzburg1964, Little1970} have existed for a long time, in recent years modifications to boson-induced superconductivity came into  focus in cold atoms \cite{Enss2009} and 2D materials \cite{Cotlet2016}.  In transition metal dichalcogenides (TMDs) strongly bound excitons~\cite{Wang2018,massignan2025} in the condensed phase play the role of bosons, mediating the interaction between fermions, similar to phonon exchange in  BCS theory~\cite{Laussy2010, Cotlet2016, Sun2021,Kumar2025}. Early quantitative investigations have focused on induced superconductivity in analogy to phonon exchange~\cite{Enss2009, Cotlet2016, Laussy2010}. Later, it  was realized that bound state physics between bosons and fermions requires to go beyond  such conventional models~\cite{Milczewski2024, Zebra2024, Zebra2025}. It was found that these strongly-coupled systems can induce superconductivity at high dimensionless  temperatures $T_c/T_F$, accompanied by an emergent BCS-BEC crossover  in which  the Fermi temperature $T_F$ sets a new bound on $T_c$~\cite{Crepel2023,Milczewski2024}. In this case, the presence of the coherence in the Bose-Einstein condensate (BEC) of bosons played an essential role, guaranteeing strong mediated forces in the fermionic system. However, one may ask what happens if the bosons become thermal and the thermal de Broglie wavelength becomes comparable to the size of Cooper pairs. Importantly, if the thermal bosons are coupled to an already established s-wave superconductor, will they disrupt the formation of Cooper pairs \cite{Antonenko2020, Smith2000} or enhance their stability \cite{Slebarski2020, Romer2018, Leroux2019}?

\begin{figure} [b] 
\centering
\includegraphics[width=0.95\columnwidth] {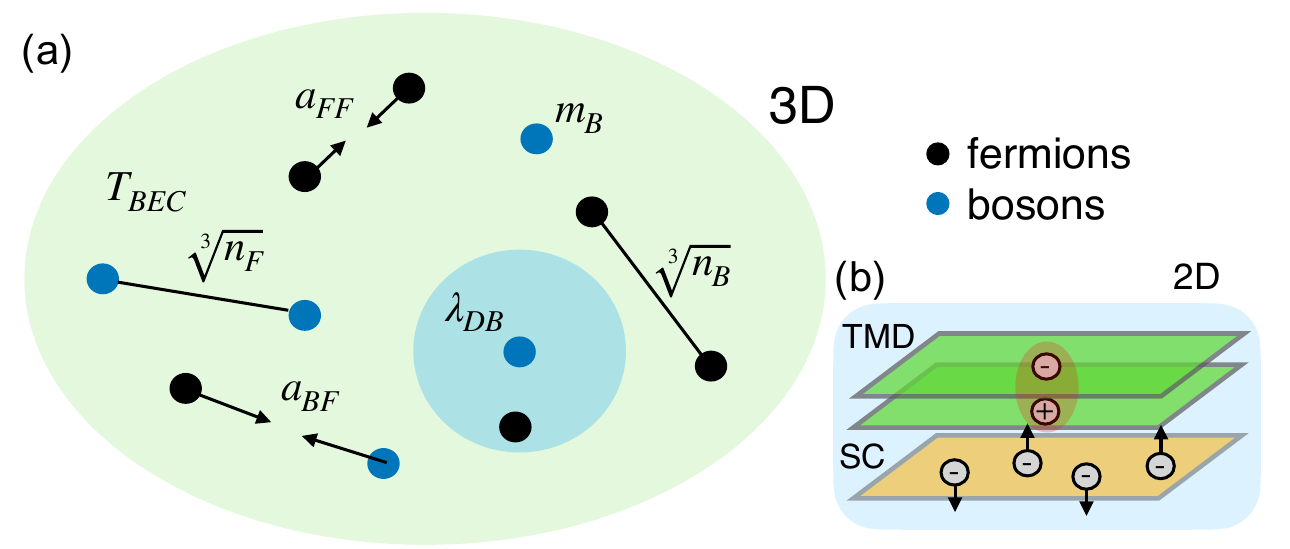}
\caption{\textbf{Platforms for thermal-boson enhanced superconductivity.} a) Ultracold atomic Bose-Fermi mixture in three dimensions. Black (blue) dots depict fermions (bosons), and  relevant
scales of the model are illustrated.  b) 2D material heterostructure where interlayer excitons in a transition metal dichalcogenide (TMD) bilayer interact with electrons in a  superconducting layer, allowing the formation of an interlayer trion (exciton-electron) bound state.
}
\label{model}
\end{figure}

In this letter, we tackle these questions using a Functional Renormalization Group (FRG) approach~\cite{Aoki2000, Bagnuls2000, Salmhofer2001, Berges2000, Pawlowski2005, Metzner2011, Wetterich_1993, Gies2012}. This allows us to consider the regime of non-perturbative boson-fermion interactions, where the absence of a clear scale separation poses a  theoretical challenge, requiring  to handle fermion-fermion  and boson-fermion interactions on equal footing. We find that accounting for the competition and interplay between bosonic and fermionic degrees of freedom is key to understand the enhancement of $T_c$ due to thermal bosons.  We perform the calculations assuming a BCS state for the underlying superconductor,  compare the critical temperature with and without the addition of bosons, and we establish a phase diagram of enhanced superconductivity for bosons being in either a thermal or condensed state.

\textbf{Model.---}Having in mind experimental realizations in both cold atoms and van der Waals materials, we consider  two-component fermions $\hat{c}_{\sigma \mathbf{k}}^{\dagger}$ of spin $\sigma$ and momentum $\mathbf{k}$, interacting with themselves and with bosonic particles $\hat{b}_{\mathbf{k}}^{\dagger}$  via attractive, contact interactions of strength $g$  and $\lambda$, respectively (see Fig.~\ref{model}). We assume no interaction between the thermal  bosons, resulting in the Hamiltonian ($\sigma=\uparrow, \downarrow$, units $\hbar = k_B=1$)
\begin{align}\label{Eq_Hamiltonian}
  H &=\sum_{\sigma,\mathbf{k}}\varepsilon_{\mathbf{k}}^c\hat{c}_{\sigma \mathbf{k}}^{\dagger}\hat{c}_{\sigma \mathbf{k}}+\sum_{\mathbf{k}}\varepsilon_{\mathbf{k}}^b\hat{b}^{\dagger}_{\mathbf{k}}\hat{b}_{\mathbf{k}} \\
 & 
+
\sum_{\mathbf{p}\mathbf{k}\mathbf{q}
} \left[g
\hat{c}^{\dagger}_{\uparrow \mathbf{p}+\mathbf{q}}\hat{c}^{\dagger}_{\downarrow \mathbf{k}-\mathbf{q}}\hat{c}_{\downarrow \mathbf{k}}\hat{c}_{\uparrow \mathbf{p}} + \sum_{\sigma}\lambda^{\sigma}\hat{c}^{\dagger}_{\sigma \mathbf{p}+\mathbf{q}}\hat{b}^{\dagger}_{\mathbf{k}-\mathbf{q}}\hat{b}_{\mathbf{k}}\hat{c}_{\sigma \mathbf{p}}\right],\nonumber
\end{align}
with $\varepsilon_{\mathbf{k}}^{c,b} = \mathbf{k}^2/2m_{F,B}-\mu_{F,B}$. 
We consider SU(2)  fermions with $m^{\sigma}_F\equiv m_F$; $\mu_F$ and $\mu_B$ are the chemical potentials of fermions and bosons,  determined by the corresponding densities $n_F$ and $n_B$ at given temperature $T$. 
In absence of the coupling to the bosons, Eq.~\eqref{Eq_Hamiltonian} gives rise to the BCS-BEC crossover,  as $g<0$ is varied from weak to strong coupling \cite{Parish, Randeria2014,zwerger2011bcs}. The coupling $\lambda^\sigma<0$ leads to a modification of $T_c$ due to the presence of bosons. While in the following we focus on the three-dimensional setting, the extension of our approach to two dimensions is straightforward, and we expect such an analysis to yield qualitatively similar results.

In this work we study how the BCS state is modified by the presence of thermal bosons. Already in the simplifying BCS limit, the system is governed by an interplay of various scales (Fig.~\ref{model}a) that feature no clear scale separation: a) the interparticle distances set by the  densities  $n_F$ and  $n_B$; b) the scattering length between fermions $a_{FF}$, and between bosons and fermions $a_{BF}$; and 
c) the thermal de Broglie wavelength $\lambda_{DB}$ beyond which bosons loose their quantum coherence. All these scales are related by  the particles' masses $m_F$ and $m_B$. 
For concreteness, in most numerical results we focus on the mass ratio $m_B/m_F=2.175$ realized in a $\text{K}^{40}$-$\text{Rb}^{87}$ Bose-Fermi mixture, and which is also close to typical mass ratios in  exciton-electron mixtures.

\textbf{Method.---}In order to analyze a system lacking scale separation, we take advantage of the functional RG (FRG). In contrast to many
functional approaches \cite{Nozieres1986, Randeria1990}, the FRG allows one to incorporate bosonic and fermionic  fluctuations on equal footing.  The central object of the  FRG is the Wetterich flow equation~\cite{Wetterich_1993}
\begin{align} \label{WetterlichEq}
    \partial_k \Gamma_{k} = \frac{1}{2} \text{STr}\left(\frac{1}{\Gamma^{(2)}_k+R_{k}}\partial_{k} R_{k}\right)\:.
\end{align}
Here $k$ is the RG scale, and the effective flowing action $\Gamma_k$  is a functional of the fields of the theory. Its  second order functional derivative  $\Gamma^{(2)}_k$ is regulated by the function $R_k$. The symbol $\rm STr$ denotes the summation over  momenta $\mathbf{p}$, Matsubara frequencies $\omega_n$, and all fields with the appropriate minus signs for  fermions.

\begin{figure*} [thb]
\centering
\includegraphics[width=0.95\textwidth] {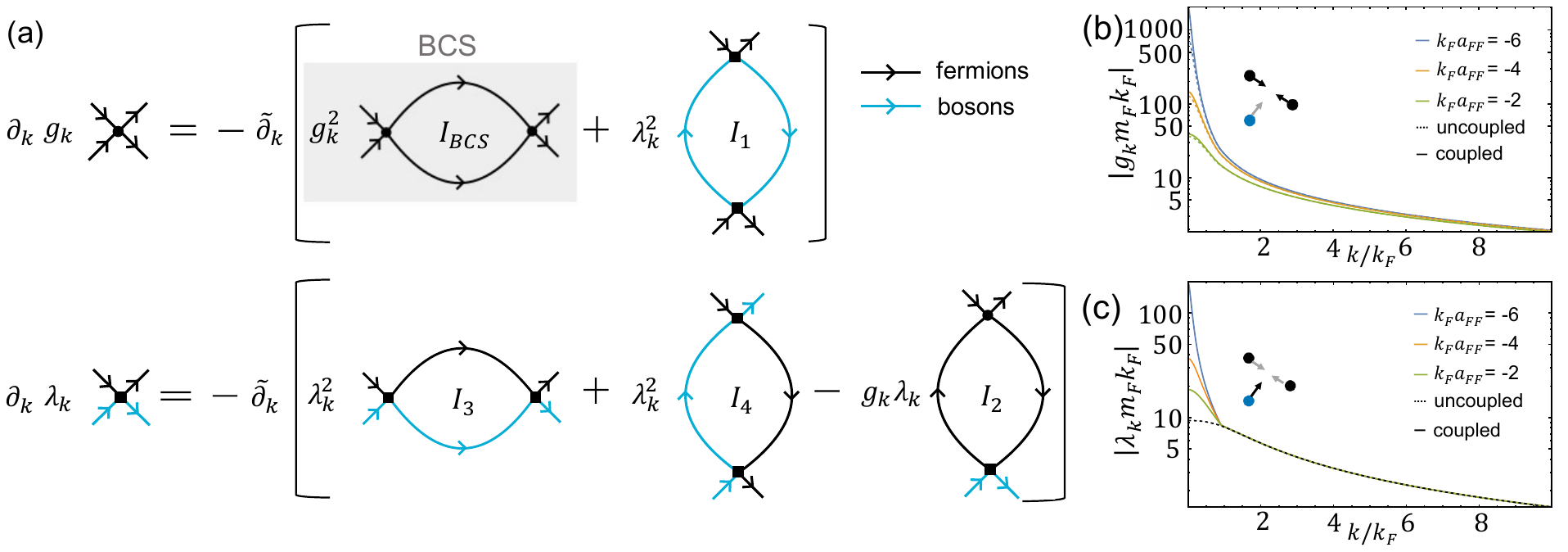}
\caption{\textbf{Renormalization group analysis.} (a) Diagrammatic representation of the RG  equations. Black (blue) lines denote fermionic (bosonic) propagators. (b,c)  RG flow with RG scale $k$ of the coupling constants $g_k$ and $\lambda_k$ with $n_B/n_F=0.1$, $(k_Fa_{BF})^{-1}=-0.2 $, $T/T_F=0.4$, $m_B/m_F=2.175$. Solid curves represent the self-consistent solution, while dashed curves show the non-selfconsistent flow ($I_1=0$ and $I_2=0$ in (a)).}
\label{FlowEquations}
\end{figure*}

In order to establish the basis for our  analysis, we briefly discuss the critical behaviour of the isolated interacting fermionic gas, governed by the BCS-BEC crossover.   BCS theory~\cite{Bardeen1957} within Leggett's approach provides a useful approximation of the crossover up to moderate interaction strength~\cite{Leggett1980, Eagles1969}. In order to reproduce the corresponding critical temperature using the FRG~\cite{schmidt2013}, we utilize the truncation of the effective action
\begin{equation}
    \begin{gathered}
    \Gamma_{k}^{BCS}=- \sum_{\sigma = \uparrow, \downarrow}\int_P\psi^*_{\sigma}(P)\left(i\omega_n -\frac{\mathbf{p}^2}{2m_F}+\mu_F\right)\psi_{\sigma}(P)\\
    +g_{k}\int_{QPP'}\psi^*_{\uparrow}(P+Q)\psi^*_{\downarrow}(P'-Q)\psi_{\downarrow}(P')\psi_{\uparrow}(P)\:,\\        
    \end{gathered}
\end{equation}
with $\int_P = T\sum_{i\omega_n}\int d^3\mathbf{p}$ and $P=(\omega_n,\mathbf{p})$ and fermionic fields $\psi$. Using sharp momentum regulators, the regularized fermionic Green's function $\left(G^c_{F}\right)^{-1} =\left(G^c_{F}\right)^{-1}+R_{F}$ takes the form $ G^c_F(\mathbf{p},\omega)= G_F(\mathbf{p},\omega)\theta\left(\left|\mathbf{p}^2/2m_F-\mu_F\right|-k^2/2m_F\right)$.
From Eq.~\eqref{WetterlichEq} one obtains the RG equation $\partial_{k} g_{k}=-g_{k}^2  \tilde{\partial}_{k}\int_Q G^c_F(Q)G^c_F(-Q)$ 
diagrammatically represented in Fig.~\ref{FlowEquations}(a). The derivative $\tilde{\partial}_{k}$ acts only on the regulator, i.e. $\tilde{\partial}_{k}= \left(\partial_{k} R_{k}\right)\partial_{R_{k}}$; for details see~\cite{SM}.

Having obtained the RG flow for the BCS state, we next include the coupling to the bosonic sector represented by the field $\phi$. To this end, we use the truncation 
\begin{equation}
\begin{gathered} \label{truncation}
    \Gamma_{k}= \Gamma_{k}^{BCS}
    -\int_P\phi^*(P)\left(i\omega_n -\frac{\mathbf{p}^2}{2m_B}+\mu_B\right)\phi(P)\\
+\sum_{\sigma}\lambda^{\sigma}_{k}\int_{QPP'} \psi^*_{\sigma}(P+Q)\phi^*(P'-Q)\phi(P')\psi_{\sigma}(P)\:,
\end{gathered}
\end{equation}
as a minimal model that captures the essential physics using flowing coupling constants $g_{{k}}$ and $\lambda^{\sigma}_{{k}}$. For example, this truncation does not account for self-energy corrections. However, for density-imbalanced systems featuring thermal bosons their effect is expected to be subleading as also recently observed in  cold atom experiments \cite{Yan2019}.

\begin{figure} [t]
\centering
\includegraphics[width=\columnwidth] {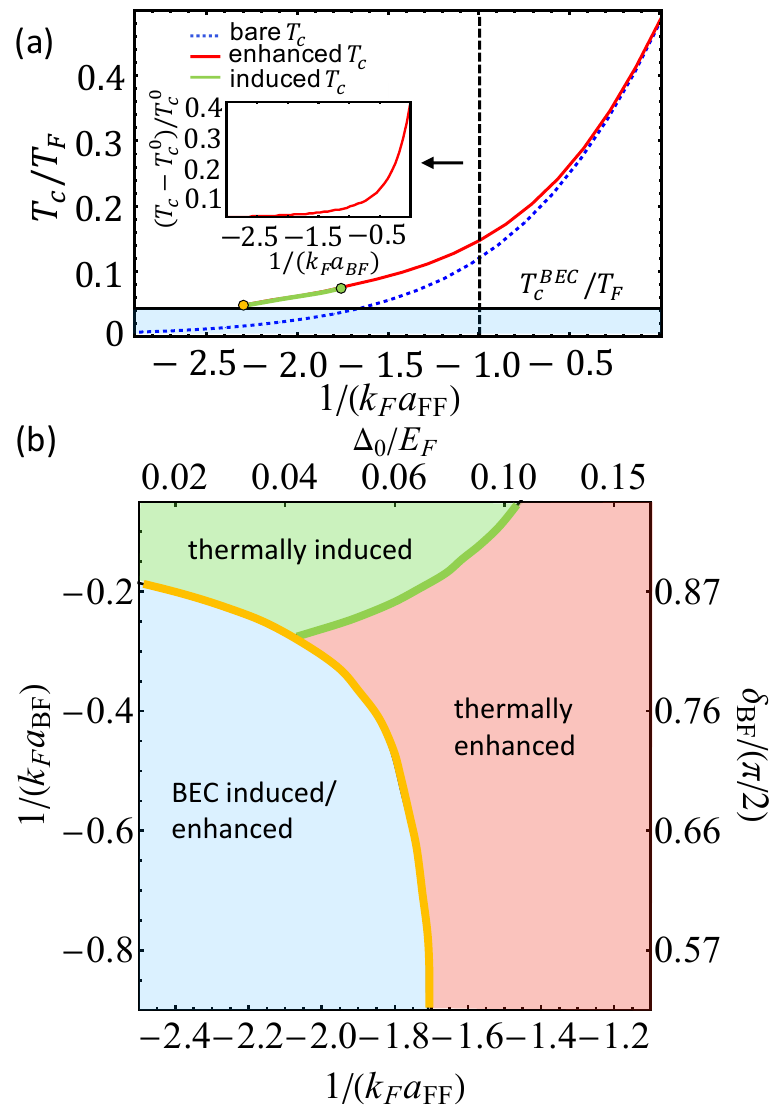}
\caption{\textbf{Thermal-boson enhanced superconductivity.}  (a) Critical temperature $T_c/T_F$ 
as function of 
$1/(k_Fa_{FF})$ at fixed boson density (solid green and red), compared to the case where bosons are absent (dashed blue). The black line shows the  temperature of  Bose-Einstein condensation $T_c^{BEC}$. Parameters are $(k_F a_{BF})^{-1}=-0.2$, $n_B/n_F=0.1$, $m_B/m_F=2.175$.  Inset: Dependence of  $T_c$ on $(k_Fa_{BF})^{-1}$ at $(k_Fa_{FF})^{-1}=-1$. (b)\:Phase diagram of  boson-modified superconductivity for a Bose-Fermi mixture  as in (a). The border between the thermally induced and enhanced superconductivity  is determined by the condition $T_c=2T_c^0$.}
\label{Tcritical}
\end{figure}

The fully interacting many-body system is described by the coupled set of RG equations~\cite{BosonRegulator} (see Fig.~\ref{FlowEquations}(a))
\begin{align} \label{FlowEq1}
        \partial_{{k}} g_{{k}} &= -g^2_{{k}} I_{BCS}
        -\lambda_{k}^{\uparrow}\lambda_{{k}}^{\downarrow}I_1 \:,\\
    \partial_{{k}} \lambda_{{k}} ^{\sigma}&= - \lambda_{{k}}^{\sigma 2}\left(I_3+I_4\right)
        +g_k\lambda_{{k}}^{-\sigma}I_2\:,
\end{align}
where $-\sigma$ denotes the opposite spin of $\sigma$. The expressions for $I_1$,...,$I_4$, and $I_{BCS}$ depend on  temperature $T$; for their explicit form see Ref.~\cite{SM}.
In the following, we assume $\lambda^{\uparrow}=\lambda^{\downarrow}=\lambda$. As in the BCS analysis, we take into account that the relevant physics occurs at the Fermi surface by projection of  momenta onto the Fermi surface at zero center-of-mass momentum. The initial conditions in the UV limit are set by the solution of the Lippmann-Schwinger equation \cite{SM}. 

The critical SC temperature $T_c$ is  determined by the divergence of the vertex $g_{k}$ in the IR limit at $k=0$, i.e. we employ the Thouless criterion $g^{-1}_{k=0}|_{T=T_c}=0$.  Numerical results for the RG flow of the vertex functions 
are shown  Fig.~\ref{FlowEquations}(b,c) at $T>T_c$, for different choices of $(k_Fa_{FF})^{-1}$. Comparing the flow of the vertices $g$, $\lambda$  in the absence and presence of the coupling between the two RG equations, one observes that the system is governed by the mutual influence of Fermi-Fermi (FF) and Bose-Fermi (BF) correlations. Indeed, due to the coupling with bosons the flow of the pairing vertex $g$ is  modified in a non-linear way by the second term in Eq.~\eqref{FlowEq1}. Similarly, the boson-fermion pairing vertices $\lambda$ is modified by the fermionic sector. The associated terms $I_1$ and $I_2$  describe density fluctuations, and we find that the  coupling between bosons and fermions enhances the attractive correlations between fermions, and between bosons and fermions alike (see illustrations in Fig.~\ref{FlowEquations}(b,c)). 

We have verified the  attractive nature of the induced interactions by comparing our result to a weak-coupling effective theory obtained by integrating out the bosons~\cite{SM}. Moreover, we find that the diagram $I_4$ ---which represents particle exchange and acts analogously to  Gorkov corrections--- only  results in a small modification of  interactions.  We note that a leading divergence of $\lambda_k$, which would signal the formation of fermionic molecules comprised of a boson and a fermion  as the new ground state of the system (i.e. trions in the context of 2D materials), does not occur. SC is thus indeed the dominating instability in our model.

\textbf{Results.---}In Fig.~\ref{Tcritical}(a) the behavior of $T_c$ is shown in  absence (dashed blue) and presence of bosons (red and green solid) at fixed density $n_B$. Answering the first question raised  in this work, we find that the critical temperature  significantly increases when thermal bosons are added to the system. This increase of $T_c$ is robust across the whole range of fermionic scattering length $a_{FF}$. The enhancement of $T_c$ by thermal bosons is, however, suppressed as the original superconductor approaches its own bound on $T_c/T_F$ close to the unitary limit. This finding is crucial as it implies that  the bound $T_c/T_F \approx 0.16$~\cite{Burovski2006, Burovski2008, Goulko2010} in the BEC-BCS crossover cannot be exceeded by coupling  fermions to a thermal bosonic bath, answering  the second question outlined in the beginning of this work.

 Analyzing results as in Fig.~\ref{Tcritical}(a) we can  construct the `phase diagram' of boson-modified superconductivity  shown in Fig.~\ref{Tcritical}(b).  We parametrize the phase diagram using two equivalent measures, chosen for easy comparison with experiments in cold atoms and 2D materials \footnote{Based on  findings on polaron physics and induced superconductivity in cold atoms and 2D materials \cite{massignan2025,Milczewski2024}, we expect our  results to qualitatively apply also for 2D systems.}. The FF interaction strength is, e.g., expressed in terms of $(k_F a_{FF})^{-1}$ known from Feshbach resonance analysis in cold atoms~\cite{Chin2010}, as well as the zero-temperature SC gap $\Delta_0$ which is  readily measurable in solid-state systems. Similarly, we give the BF interaction strength as function of $(k_F a_{BF})^{-1}$ as well as the scattering phase shift $\delta_{BF}$ \footnote{The BCS weak-coupling gap and the phase-shift are calculated using the  relations $\Delta_0=1.764 T_c^0$ and $\delta_{BF}=\text{arccot}(-1/(k_Fa_{BF}))$.}, which has been characterized for exciton-electron mixtures  and can be readily connected to trion binding energies \cite{Fey2020,Levinsen2025,Wagner2025,Deilmann2026}.

\begin{figure}[b] 
\centering
\includegraphics[width=0.95\columnwidth] {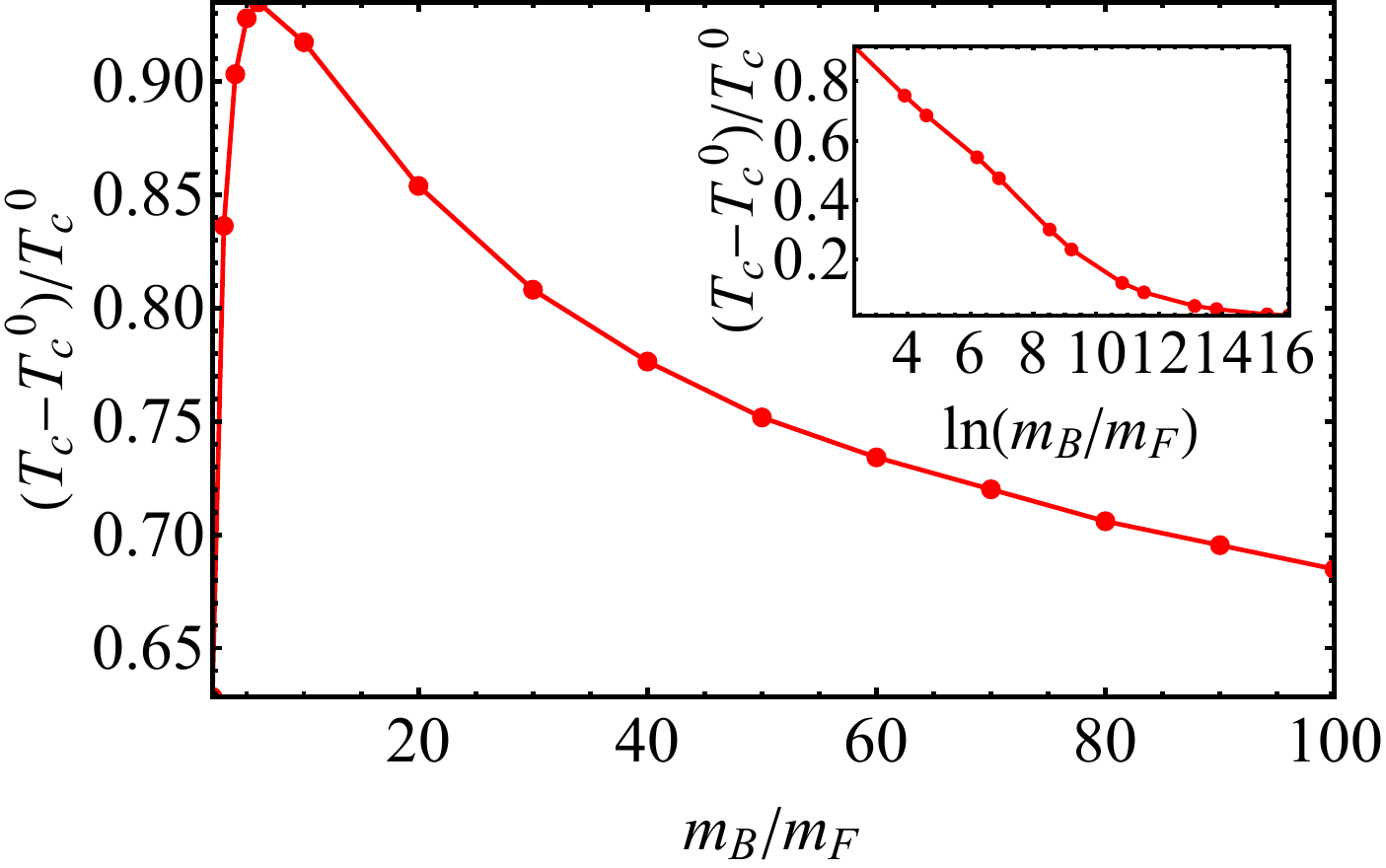}
\caption{\textbf{Mass scaling.}  Enhancement of  $T_c$ relative to its bare value $T_c^0$ as function of the mass ratio $m_B/m_F$ for $n_B/n_F=0.1 $, $(k_F a_{BF})^{-1}=-0.2$ and $(k_F a_{FF})^{-1}=-1.5$. Inset: Log-linear plot showing $T_c$ up to large values of $m_B/m_F$.}
\label{MassDependence}
\end{figure}

We identify three phases as function of the Bose-Fermi and Fermi-Fermi interaction strengths. First, the weak coupling regime is governed by a phase which requires bosons to be in the superfluid state to enhance SC (shaded blue). Here the system only turns superconducting at temperatures below $T_c^{BEC}$ of Bose condensation,  i.e. for $T<T_c^{BEC}\approx 3.3\: n_B^{2/3}/m_B$ for an ideal Bose gas in 3D (black solid line in Fig.~\ref{Tcritical}(a)).

At strong coupling $1/(k_F a_{FF})\gtrsim -1.7$, but intermediate values of $k_Fa_{BF}$, the modified $T_c$ raises above $T_c^{BEC}$. Here thermal bosons are sufficient to enhance SC (shaded red in~Fig.~\ref{Tcritical}(b)). At large BF interactions, we find that the thermal bosons provide the dominant mechanism driving SC. This results in an increase of $T_c$ by more than a factor of two, $T_c/T_c^{0}>2$. Reflecting this dominance of BF coupling, we term this regime as the phase of thermal-boson induced superconductivity. Naturally, the  magnitude of $n_B$   influences the enhancement of $T_c$. Driven by the last term in Eq.~\eqref{FlowEq1}, $T_c$  scales approximately linear with $n_B$; corresponding phase diagrams for different densities are presented in Ref.~\cite{SM}.

The boson mass is a further central parameter in determining the enhancement of $T_c$. Naively one might expect that a lighter mediator of  interactions may be beneficial. For thermal-boson enhanced SC we find this not to be the case. As shown in Fig.~\ref{MassDependence},  making the bosons heavier first  enhances mediated interactions. Only for large boson mass the enhancement weakens and eventually  recovers the original value of $T_c^0$.  Interestingly, a similar behavior of mixed boson-fermion systems was recently found in a study of the Fermi polaron problem where  a significant  mass imbalance leads to enhanced polaron dressing~\cite{Chen2025}. Attesting to the strength of the FRG, we note that without the self-consistent feedback between   $\lambda$  and $g$  (Fig.~\ref{FlowEquations}(a)) one would obtain a monotonous, non-physical increase of the critical temperature extending into the limit $m_B/m_F\rightarrow\infty$~\cite{SM}. We also note that we do not include a polaronic renormalization of the fermionic mass in our analysis.  However, as we consider a thermal bath such self-energy effects are expected to be small.  Moreover, as fermions become heavier, the associated increase in the density of state may even further enhance $T_c$.

We have focused the discussion on the SC instability. To support our findings, we finally investigate the robustness of the SC transition with regards to a charge density wave (CDW) instability. Indeed, since $g$ and $\lambda$ correspond to effective in-medium scattering amplitudes, their mutual increase might hint at a tendency towards enhanced density-density correlations. To analyze the role of a potential CDW instability, we couple the action to source fields~\cite{Halboth2000, Salmhofer2001, Metzner2012,Desoppi2024} and monitor the  divergence of associated vertex functions~\cite{SM}. We find that within our truncation  superconductivity always dominates. This finding is also in line with the fact that the boson dispersion  in our model does not feature  softening at a certain wavevector, which was a crucial ingredient to trigger CDW instabilities in previous work \cite{Enss2009,Cotlet2016}.

Experimentally, our model can be realized in ultracold atomic  mixtures featuring overlapping Feshbach resonances~\cite{Goldwin2004, Roati2002}. The critical temperature can be measured in rotating paired superfluids~\cite{Zwierlein2005} or by spectroscopy of the pairing gap~\cite{Chin2004, Stewart2008}. The predicted enhancement of correlations  can  be observed using atom-resolved techniques in continuum \cite{Yao2025, deJongh2025, Xiang2025}, and  complementary momentum-space techniques \cite{Jeltes2007}. In 2D materials, transport measurements are available as a direct probe of boson-enhanced superconductivity. Interestingly, the measurement of thermal-boson enhanced SC may provide a novel probe to observe the transition of excitonic insulators from the normal to the  superfluid regime. Possible setups could include heterobilayers of TMDs coupled to a superconducting layer of (twisted) bilayer graphene which is in reach of current experimental techniques~\cite{Cao2018, Pasqual2016,Inbar2023,Ma2021}.

\textbf{Conclusion.---}We have shown that the coupling of a superconductor to a thermal boson bath can lead to a robust increase of the SC temperature $T_c$.  In this way, $T_c$ can reach more than twice of its bare value at moderate fermion coupling. For strongly-coupled fermion system,  the enhancement becomes increasingly suppressed.  As a result, the bound on $T_c/T_F\approx0.16$  for the unitary Fermi gas remains robust.  While we have mainly focused on  three dimensions, we expect our results to be qualitatively applicable to the study of two-dimensional van der Waals materials. In this context, the extension of the present theory to include the  physics of the Berezinskii–Kosterlitz–Thouless phase transition may allow to establish thermally-enhanced SC as a novel probe of the physics of excitonic insulators. A further interesting theoretical venue will be the full incorporation of three-body correlations and three-body loss which will be of particular importance in the regime of small boson masses which has not been focus of this work. Finally, the incorporation of a full BCS-BEC crossover \cite{Milczewski2024,Crepel2023} will require the consideration of tightly bound fermions pairs  which presents another interesting direction  for future work.

\textbf{Acknowledgements.---}We thank  Xin Chen, Jonas von Milczewski and Atac Imamoglu for inspiring discussions. E.~V., M.~S., and R.~S. acknowledge support by the Deutsche Forschungsgemeinschaft under Germany's Excellence Strategy EXC 2181/1 - 390900948 (the Heidelberg STRUCTURES Excellence Cluster), and CRC 1225 ISOQUANT, project-ID 273811115. R.S. was supported within the DFG scientific network `A(E)MP - Appearance of the Effective Mass in Polaron Models' (Grant No. 569490025). E.~D. acknowledges support from the Swiss National Science Foundation (SNSF) under Project 200021\_212899 and  Sinergia SNF\_222792.


\begin{thebibliography}{75}
\bibitem [{Bos()}]{BosonRegulator}%
  \BibitemOpen
  \bibinfo {note} {Analogously to the fermions the bosonic Green's function is regulated by a sharp regulator $G^c_B(\mathbf{p},\omega)=G_B(\mathbf{p},\omega)\theta \left(p-k\right)$.}\BibitemShut {Stop}%
\end{thebibliography}
%

\end{document}


\title{Supplemental Material for \\ `Enhancing superconductivity using thermal bosons'} 

\author{Ekaterina Vlasiuk}
\affiliation{Institute for Theoretical Physics, Heidelberg University, 69120 Heidelberg, Germany}

\author{Manfred Salmhofer}
\affiliation{Institute for Theoretical Physics, Heidelberg University, 69120 Heidelberg, Germany}

\author{Eugene Demler}
\affiliation{Institute for Theoretical Physics, ETH Z\"urich, 8093 Z\"urich, Switzerland}

\author{Richard Schmidt}
\affiliation{Institute for Theoretical Physics, Heidelberg University, 69120 Heidelberg, Germany}

\maketitle

In this supplemental material we provide further information and details on our calculations leading to the results presented in the main text of our work. Specifically,  in Sec.~\ref{sec:BCS} we show how the well-known BCS results are obtained within the Functional Renormalization Group (FRG) framework, and discuss the Gorkov-Melik-Barkhudarov correction to BCS theory. 
In Secs.~\ref{sec:IntExpr} and~\ref{sec:InitCond} we present the integral expression entering the RG flow equations for the Bose-Fermi mixture, and we discuss the associated initial conditions. In Sec.~\ref{sec:FixedPoints} we present the fixed point diagram in the vacuum limit, while in Sec.~\ref{sec:EffTheory} we discuss the derivation of an effective weak-coupling theory for the fermions following integrating out the bosonic degrees of freedom.  In Sec.~\ref{sec:PhaseDiagrams} we present  phase diagrams of  enhanced/induced superconductivity in  Bose-Fermi mixtures for various density ratios between bosons and fermions. In Sec.~\ref{sec:NonphysicalBehaviour} we discuss the nonphysical behaviour of the critical superconducting temperature when the problem is solved in a non-self-consistent approach. Finally, in Sec.~\ref{sec:ChargeDensity} we discuss and analyse the role  of a potential charge density wave instability.

\section{BCS theory from the functional renormalization group and Gorkov-Melik-Barkhudarov (GBM) correction}\label{sec:BCS}
The well-known results from BCS theory can be obtained from the Functional Renormalization Group (FRG) for the purely fermionic system by keeping only the first diagram in Fig.~2(a) of the main text. This leads to the  flow equation 
\begin{equation}
    \partial_{k} g_{k}=-g_{k}^2  \tilde{\partial}_{k}\int_Q G^c_F(Q)G^c_F(-Q)\:.
\end{equation}

In the weak coupling limit, $k_F a\rightarrow 0^-$, one then  obtains~\cite{schmidt2013}
\begin{equation}
    T_c^0=\frac{8\e^{\gamma}}{\pi e^2}e^{-\frac{\pi}{2}\left|\frac{1}{k_Fa}\right|}\:,
\end{equation}
recovering the BCS result.  In the unitary limit $1/k_F a\rightarrow 0$ one arrives at at the well-known  result $T_c^0/E_F\approx0.5$ (see the dashed blue curve in Fig.~3(a)) \cite{SaDeMelo1993}, which can  also be obtained using a non-self-consistent T-matrix approach.

\begin{figure} [h]
\centering
\includegraphics[width=12cm] {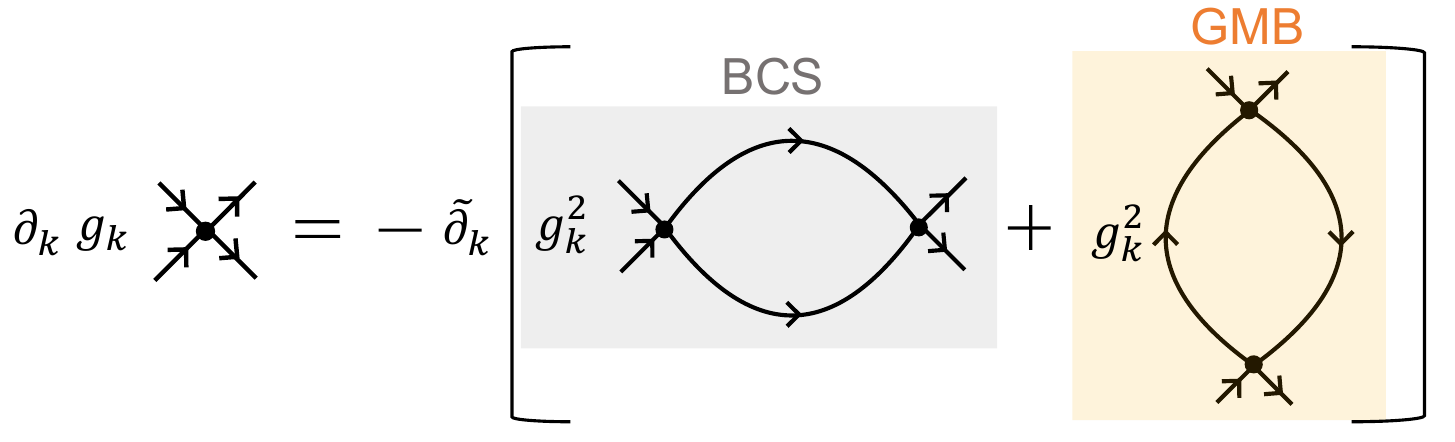}
\caption{Flow equation for the flowing coupling constant between fermions $g_k$ in the pure fermionic theory with the inclusion of Gorkov-Melik-Barkhudarov correction.}
\label{GMB}
\end{figure}

For the transparency and simple analysis of our results we neglect the Gorkov-Melik-Barkhudarov (GMB) correction in the main text. One generally obtains the GMB correction following the FRG procedure outlined in the main text. The associated  flow equation for the pure fermionic model including the GMB correction (for its diagrammatic representation, see Fig.~\ref{GMB}) is given by
\begin{equation}
    \partial_{\text{k}} g_{\text{k}} = -g^2_{\text{k}} I_{BCS}
        -g_{\text{k}}^2 I_2\:.
\end{equation}
The GMB correction would  overall  lower the  critical temperature in comparison to  pure BCS theory but it would not qualitatively impact the mechanism leading to the enhancement of $T_c$ by thermal bosons. 

\section{Integral expressions}\label{sec:IntExpr}
Here we provide the integral expressions entering the flow equations~(5)-(6) of the main text. They are given by
\begin{align}
    \begin{gathered}
        I_{BCS}=\tilde{\partial}_k\int_Q G^c_F(Q)G^c_F(-Q)= -\frac{m_F}{2\pi^2k}\tanh\left[\frac{k^2}{4m_FT}\right]\left\{\theta\left(\mu_F-\frac{k^2}{2m_F}\right)\sqrt{2m_F\mu_F-k^2}+\sqrt{2m_F\mu_F+k^2}\right\}\:,
    \end{gathered}
\end{align}
and
\begin{align}
    \begin{gathered}
            I_1=\frac{1}{2}\int_{-1}^1d\cos\varphi\:\tilde{\partial}_k\int_Q G_B^c(Q)G_B^c(Q+L_3-L_2)=-\frac{k^2 m_B}{2 \pi^2 k_F^2}\Bigg[\left(1+\frac{k_F}{k}\right)\ln\left(1+\frac{k_F}{k}\right)\\
            +\left(1-\frac{k_F}{k}\right)\ln \left|1-\frac{k_F}{k}\right|\Bigg]n_B\left(\frac{k^2}{2m_B}-\mu_B\right) \:,
    \end{gathered}
\end{align}
where $\varphi =\angle(l_2,l_3).$ Moreover, 
\begin{align}
    \begin{gathered}
        I_2=\frac{1}{2}\int_{-1}^1d\cos\varphi \:\tilde{\partial}_k \int_Q G_F^c(Q)G_F^c(Q+L_1-L_4)=\frac{m_Fk \sqrt{2m_F\mu_F+k^2}}{2\pi^2k_F^2}\\
        \times\Bigg[\left(1+\frac{k_F}{\sqrt{2m_F\mu_F+k^2}}\right)\ln \left(1+\frac{k_F}{\sqrt{2m_F\mu_F+k^2}}\right)\\
    +\left(1-\frac{k_F}{\sqrt{2m_F\mu_F+k^2}}\right)\ln \left|1-\frac{k_F}{\sqrt{2m_F\mu_F+k^2}}\right|\Bigg]n_F\left(\frac{k^2}{2m_F}\right)\\
    +\frac{m_F k \sqrt{2m_F\mu_F-k^2}}{2\pi^2k_F^2}\Bigg[\left(1+\frac{k_F}{\sqrt{2m_F\mu_F-k^2}}\right)\ln \left(1+\frac{k_F}{\sqrt{2m_F\mu_F-k^2}}\right)\\
    +\left(1-\frac{k_F}{\sqrt{2m_F\mu_F-k^2}}\right)\ln \left|1-\frac{k_F}{\sqrt{2m_F\mu_F-k^2}}\right|\Bigg]n_F\left(-\frac{k^2}{2m_F}\right)\theta\left(\mu_F-\frac{k^2}{2m_F}\right)\:,
    \end{gathered}
\end{align}
where $\varphi =\angle(l_1,l_4)$, and 
\begin{align}
    \begin{gathered}
        I_3=\int_Q G_F^c(-Q+L_4+L_3)G_B^c(Q)=\int_Q G_F^c(-Q)G_B^c(Q) \\
        = -\frac{k\sqrt{2m_F\mu_F+k^2}}{2\pi^2}\left[\frac{n_F\left(-\frac{k^2}{2m_F}\right)+n_B\left(\frac{k^2}{2m_B}+\frac{m_F}{m_B}\mu_F-\mu_B\right)}{\left(\frac{1}{2}\left(\frac{1}{m_B}+\frac{1}{m_F}\right)(2m_F\mu_F+k^2)-\mu_B-\mu_F\right)^2+\left(\pi T \right)^2}\right]\\
        \times\left[\frac{1}{2}\left(\frac{1}{m_B}+\frac{1}{m_F}\right)(2m_F\mu_F+k^2)-\mu_B-\mu_F\right]\\
        -\frac{k^2}{2\pi^2}\left[\frac{n_F\left(-\frac{k^2}{2m_F}+\mu_F\right)+n_B\left(\frac{k^2}{2m_B}-\mu_B\right)}{\left(\frac{1}{2}\left(\frac{1}{m_B}+\frac{1}{m_F}\right)k^2-\mu_B-\mu_F\right)^2+\left(\pi T\right)^2}\right]\left[\frac{1}{2}\left(\frac{1}{m_B}+\frac{1}{m_F}\right)k^2-\mu_B-\mu_F\right]\theta\left(\mu_F-\frac{k^2}{m_F}\right)
    \end{gathered}
\end{align}

\begin{align}
    \begin{gathered}
        I_4=-\frac{1}{2}\int_{-1}^1d\cos\varphi\:\int_Q G_F^c(Q+L_4-L_2)G_B^c(Q)\\
        =-\frac{1}{2}\int_{-1}^1dy\Bigg\{\Bigg[\frac{m_Bk}{4\pi^2p}n_F\left(\frac{q^2}{2m_F}-\mu_F\right)\ln\left|\frac{-\frac{qp}{m_B} +\frac{q^2}{2}\left(\frac{1}{m_B}-\frac{1}{m_F}\right) +\frac{p^2}{2m_B}+\mu_F-\mu_B}{+\frac{qp}{m_B} +\frac{q^2}{2}\left(\frac{1}{m_B}-\frac{1}{m_F}\right) +\frac{p^2}{2m_B}+\mu_F-\mu_B}\right|\\
        +\frac{m_Fk}{4\pi^2 p}n_B\left(\frac{q^2}{2m_B}-\mu_B\right)\ln\left|\frac{-\frac{qp}{m_F}+\frac{q^2}{2}\left(\frac{1}{m_B}-\frac{1}{m_F}\right)-\frac{p^2}{2m_F}+\mu_F-\mu_B}{+\frac{qp}{m_F}+\frac{q^2}{2}\left(\frac{1}{m_B}-\frac{1}{m_F}\right)-\frac{p^2}{2m_F}+\mu_F-\mu_B}\right|\Bigg] \\
        +\Bigg[\frac{m_Bk}{4\pi^2p}n_F\left(\frac{k^2}{2m_F}-\mu_F\right)\ln\left|\frac{-\frac{kp}{m_B} +\frac{k^2}{2}\left(\frac{1}{m_B}-\frac{1}{m_F}\right) +\frac{p^2}{2m_B}+\mu_F-\mu_B}{+\frac{kp}{m_B} +\frac{k^2}{2}\left(\frac{1}{m_B}-\frac{1}{m_F}\right) +\frac{p^2}{2m_B}+\mu_F-\mu_B}\right|\\
        +\frac{m_Fk}{4\pi^2 p}n_B\left(\frac{k^2}{2m_B}-\mu_B\right)\ln\left|\frac{-\frac{kp}{m_F}+\frac{k^2}{2}\left(\frac{1}{m_B}-\frac{1}{m_F}\right)-\frac{p^2}{2m_F}+\mu_F-\mu_B}{+\frac{kp}{m_F}+\frac{k^2}{2}\left(\frac{1}{m_B}-\frac{1}{m_F}\right)-\frac{p^2}{2m_F}+\mu_F-\mu_B}\right|\Bigg]\theta\left(\mu_F-\frac{k^2}{m_F}\right)\Bigg\}\:,
    \end{gathered}
\end{align}
where $q=\sqrt{2m_F\mu_F+k^2}$, $p=\sqrt{2}k_F\sqrt{1-y}$.

In these equations $L_3$ and $L_4$ denote incoming external momenta and frequencies of the scattering fermions while $L_1$ and $L_2$ are outgoing external momenta and frequencies. In the above derivation we considered the low-energy physics by projecting the external momenta onto the Fermi surface at zero center-of-mass momentum, and we assumed that the chemical potential of fermions $\mu_F$ is positive, which is a valid assumption for low enough temperatures. Similar to previous studies~\cite{Desoppi2024}, we ignored the external momentum dependence in the regulator function in favor of obtaining analytical results, as this approach does not influence the results on the qualitative level.

\section{Initial conditions}\label{sec:InitCond}
For the solution of the flow equations one has to set the initial conditions. This means that one has to find some limit in which the solution of the flow equations is known. This is achieved by considering  the solution at the UV scale, where it is fully  governed by short-distance, and thus two-body scattering physics. In the two-body limit (also called `vacuum limit'), the chemical potentials and temperature are set to zero ($\mu_F=\mu_B=0$, $T=0$). 
From the flow equations in this limit one obtains 
\begin{equation}
    \begin{gathered}
       g_{{k}=\Lambda}=\left(\frac{m_F}{4\pi a_{FF}}-\frac{m_F}{2\pi^2}\Lambda\right)^{-1}\:,\\
       \lambda_{{k}=\Lambda}=\left(\frac{1}{2\pi a_{BF}}\frac{m_Bm_F}{m_B+m_F}-\frac{1}{\pi^2}\frac{m_Bm_F}{m_B+m_F}\Lambda\right)^{-1}\:,
    \end{gathered}
\end{equation}
where we used that in the two-body limit $g_{{k}=0}=4\pi a_{FF}/\mu_{FF}$ and $\lambda_{{k}=0}=2\pi a_{BF}(m_B+m_F)/(m_Bm_F)$ as the flowing constants are the zero-energy two-body T-matrices which in its turn are connected to the scattering amplitudes and the scattering wavelengths.  
We utilize these expressions as the initial conditions for the flow equations.

\section{Fixed points diagram in the vacuum  
limit}\label{sec:FixedPoints}

\begin{figure} [thb]
\centering
\includegraphics[width=7cm] {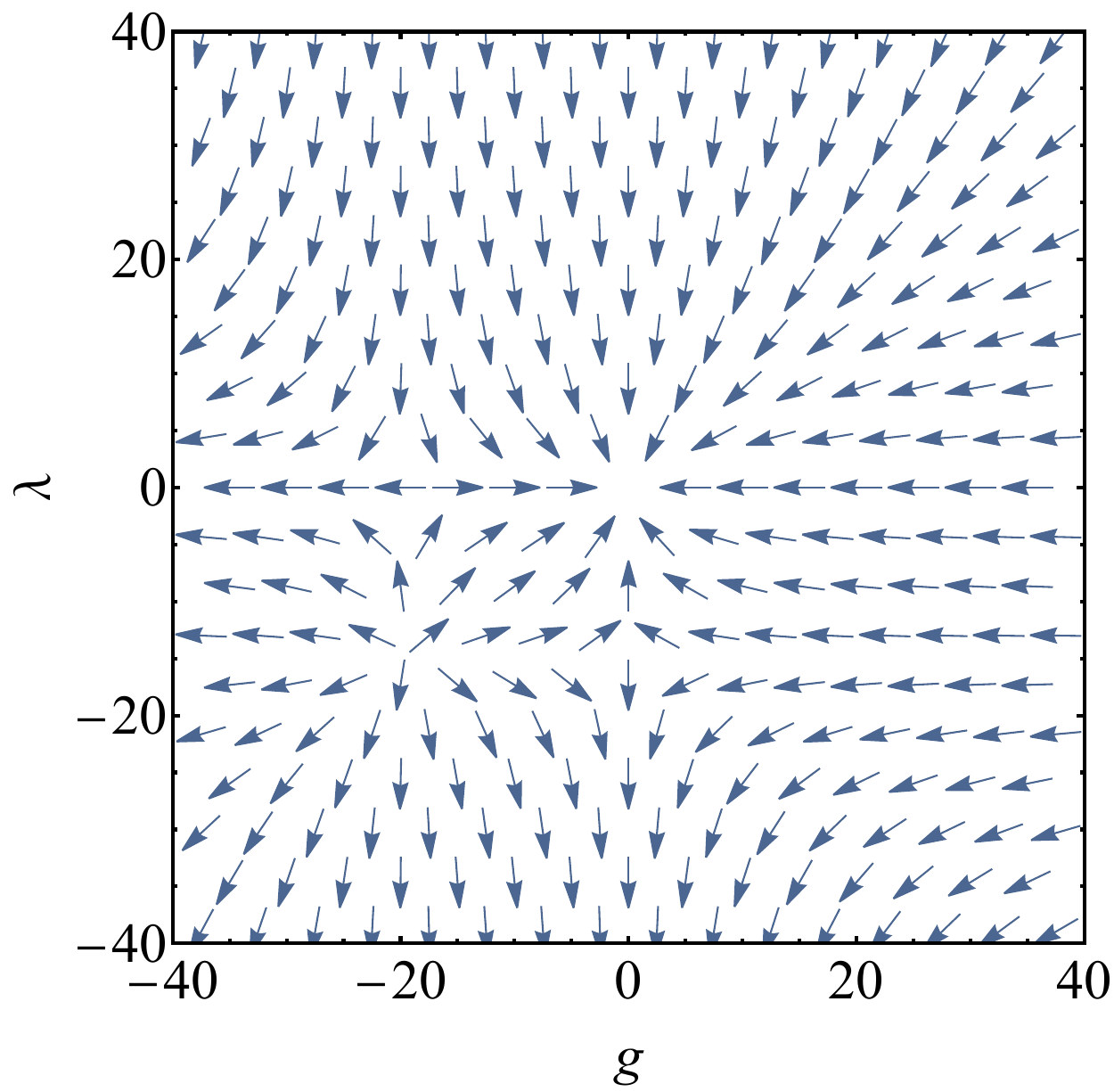}
\caption{Fixed points diagram of the RG flow equations in the vacuum limit for $\text{K}^{40}$-$\text{Rb}^{87}$ mixture. The arrows show the direction towards the IR limit.}\label{figRGDiag}
\end{figure} 

The tendencies in the behaviour of the flowing coupling constants can be already seen in the vacuum limit at vanishing densities and temperature. Here the flow equations are given by
\begin{equation}
    \begin{gathered}
    \partial_k \tilde{g}_k = \frac{m_F}{2\pi^2} \tilde{g}_k ^2 + \tilde{g}_k\:,\\
    \partial_k \tilde{\lambda}_k = \frac{1}{\pi^2}\frac{m_B m_F}{m_B + m_F}\tilde{\lambda}_k^2+ \tilde{\lambda}_k\:,      
    \end{gathered}
\end{equation}
where we use rescaled quantities $k=\Lambda e^t$, $\tilde{g}=g k$ and $\tilde{\lambda}=\lambda k $. The RG  diagram  in Fig.~\ref{figRGDiag} shows the strong coupling fixed points and instability towards fermion-fermion and fermion-boson pairing, i.e. $g\rightarrow -\infty$ $\lambda \rightarrow -\infty$, which drives the physics described in this work.

\section{Effective theory in the weak coupling limit}\label{sec:EffTheory}
The action is quadratic in bosonic fields. Thus it is possible to integrate the bosons out exactly, which, when performed  in the limit of the weak-coupling between bosons and fermions, leads to an analytical understanding of boson-mediated interactions. In particular it provides a solid check of the character of the induced interaction.  Of course, this approach will not allow to reach the strong-coupling regime, as keeping only lowest-order induced interactions in this approach, one cannot  account for the mutual influence between different scattering channels which is crucial, for instance, for uncovering the nontrivial mass dependence of the critical temperature discussed in the main text. 
To obtain the effective weak-coupling theory, we start with the action for the Bose-Fermi mixture
\begin{equation}
    \begin{gathered}
        S= \sum_{\sigma = \uparrow, \downarrow}\int_P\psi^*_{\sigma}(P)\left(i\omega -\frac{p^2}{2m_F}+\mu_F\right)\psi_{\sigma}(P) + \int_P\varphi^*(P)\left(i\omega -\frac{p^2}{2m_B}+\mu_B\right)\varphi(P)\\ 
        +g_{k}\int_{Q,P,P'}\psi^*_{\uparrow}(P+Q)\psi^*_{\downarrow}(P'-Q)\psi_{\downarrow}(P')\psi_{\uparrow}(P)\\
        +\sum_{\sigma=\uparrow,\downarrow}\lambda^{\sigma}\int_{Q,P,K'} \psi^*_{\sigma}(P+Q)\varphi^*(K'-Q)\varphi(K')\psi_{\sigma}(P)\:.
    \end{gathered}
\end{equation}
The partition function is then given by
\begin{equation}
    \begin{gathered}
        Z = \int D [\psi_{\uparrow},\psi_{\uparrow}^*,\psi_{\downarrow}, \psi_{\downarrow}^*, \varphi, \varphi^*] \exp\left({-S_F[\psi_{\uparrow},\psi_{\uparrow}^*,\psi_{\downarrow}, \psi_{\downarrow}^*] -\sum_{KK'}\varphi^*(K) K_{KK'}[\psi_{\uparrow},\psi_{\uparrow}^*,\psi_{\downarrow}, \psi_{\downarrow}^*]\varphi(K')}\right)\:,\\
K[\psi_{\uparrow},\psi_{\uparrow}^*,\psi_{\downarrow}, \psi_{\downarrow}^*]_{KK'}=G_B^{-1}\delta_{K,K'}+\sum_{\sigma=\uparrow,\downarrow}\lambda^{\sigma}\left(N_{F\sigma}\right)_{KK'}\textbf{, }\\
\left(N_{F\sigma}\right)_{KK'}=\int_Q n_{F\sigma}(Q)\delta_{K',K+Q}\text{, }
n_{F\sigma}(Q) =\int_P \psi_{\sigma}^*(P+Q)\psi_{\sigma}(P)\:.
\end{gathered}
\end{equation}
We now perform the functional integration, followed by the expansion of the result
\begin{equation} \label{FieldInt}
    \int D[\varphi, \varphi^*] \exp\left( -\sum_{KK'}\varphi^*(K) K_{KK'}[\psi_{\uparrow},\psi_{\uparrow}^*,\psi_{\downarrow}, \psi_{\downarrow}^*]\varphi(K')\right)=\frac{C}{\det K}\text{, }C>0\:,
\end{equation}
\begin{equation}
\begin{gathered}
    \det K = \exp \left(\text{Tr}\ln K\right)= \exp\left[ \text{Tr}\ln G_B^{-1} + \text{Tr}\ln \left(1+\sum_{\sigma=\uparrow,\downarrow}\lambda^{\sigma}N_{F\sigma}G_B\right)\right]\\
    \approx \exp\left[ \text{Tr}\ln G_B^{-1} + \text{Tr}\left(\sum_{\sigma=\uparrow,\downarrow}\lambda^{\sigma}N_{F\sigma}G_B\right) -\frac{1}{2} \text{Tr} \left(\sum_{\sigma=\uparrow,\downarrow, \sigma'=\uparrow,\downarrow}\lambda^{\sigma}\lambda^{\sigma'}N_{F\sigma}G_BN_{F\sigma'}G_B\right)\right]\:.
\end{gathered}
\end{equation}
Taking the definition of the effective action $Z=\int D [\psi_{\uparrow},\psi_{\uparrow}^*,\psi_{\downarrow}, \psi_{\downarrow}^*] \exp\left(-S_{\text{eff}}\right)$ into account we obtain
\begin{equation}
\begin{gathered}
S_{\text{eff}}=S_F[\psi_{\uparrow},\psi_{\uparrow}^*,\psi_{\downarrow}, \psi_{\downarrow}^*] +\text{Tr}\ln G_B^{-1} + \text{Tr}\left(\sum_{\sigma=\uparrow,\downarrow}\lambda^{\sigma}N_{F\sigma}G_B\right) -\frac{1}{2} \text{Tr} \left(\sum_{\sigma=\uparrow,\downarrow, \sigma'=\uparrow,\downarrow}\lambda^{\sigma}\lambda^{\sigma'}N_{F\sigma}G_BN_{F\sigma'}G_B\right) \:.
\end{gathered}
\end{equation}
One can see that the interaction term which is quadratic in $\lambda$ is attractive since 
\begin{equation} \label{f(k)}
    \begin{gathered}
         \text{Tr} \left(N_{F}G_BN_{F}G_B\right) = \int dK f(K) n_F(K)n_F(-K)\:,\:f(K)=\int dK_1G_B(K_1)G_B(K_1+K)>0\:.
    \end{gathered}
\end{equation}
This result agrees with the FRG calculation in the main text, namely, the $\lambda_k^{\uparrow}\lambda_k^{\downarrow}$ term in Eq.~(5) is  attractive.

We note that in the reverse case where the interaction between fermions is absent one can obtain an effective induced interaction between bosons by performing the same procedure as above by integrating out the fermionic fields. The difference is that in Eq.~\eqref{FieldInt} the determinant would be in the numerator, and the sign of the function analogous to $f(K)$ in Eq.~\eqref{f(k)}, but with fermionic propagators, would be opposite. As a result, the quadratic term in the effective bosonic action would also be  attractive. When extending the truncation of our FRG approach to featuring induced bosonic interactions we would thus also find induced attraction between bosons. The resulting  density collapse of the system would,  however, be prevented by a sufficiently strong two-body repulsion between bosons.

\section{Phase diagrams for different densities} \label{sec:PhaseDiagrams}
\begin{figure} [h]
\centering
\includegraphics[width=18cm] {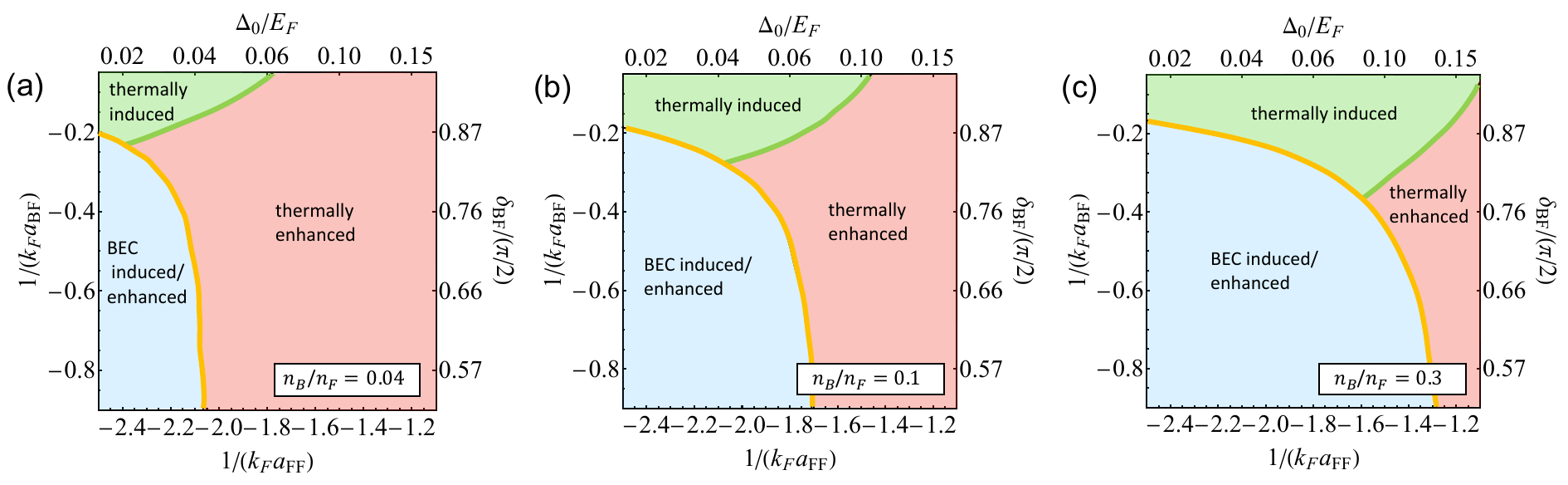}
\caption{Phase diagram of boson-modified
superconductivity for a $\text{K}^{40}$-$\text{Rb}^{87}$ Bose-Fermi mixture for the different densities: (a) $n_B/n_F=0.04$, (b) $n_B/n_F=0.1$, (c) $n_B/n_F=0.3$. The border between the thermally induced and thermally enhanced superconductivity regions is defined by the condition $T_c=2T_c^0$.} \label{densfigSM}
\end{figure}

In this section we present the  phase diagrams of  enhanced/induced superconductivity in a $\text{K}^{40}$-$\text{Rb}^{87}$ Bose-Fermi mixture for  different boson densities $n_B$. The results are shown in Fig.~\ref{densfigSM} which are obtained analogously to the results shown in the main text.

\section{Nonphysical behaviour of the critical temperature within the non-self-consistent solution of the flow equations} \label{sec:NonphysicalBehaviour}

\begin{figure} [t!]
\centering
\includegraphics[width=8cm] {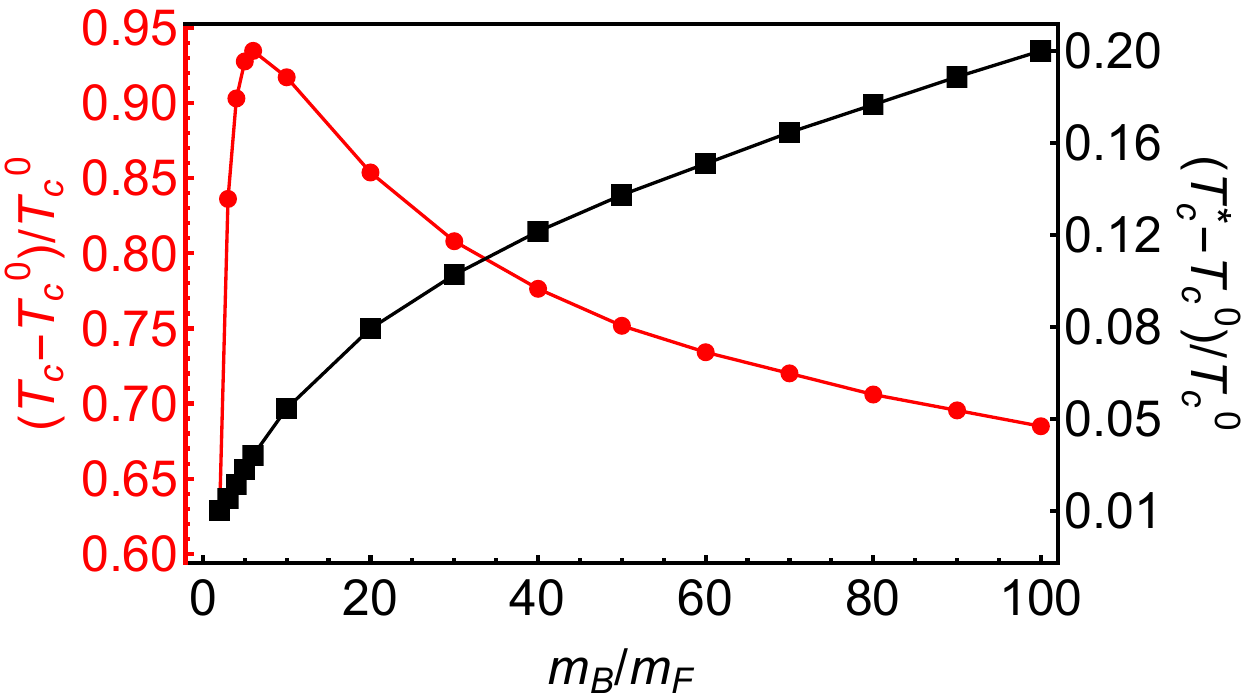}
\caption{The dependence of the critical superconducting temperature $T_c$ on the mass ratio between bosons and fermions $m_B/m_F$ with (red curve) and without (black curve) having a feedback of $\lambda$-vertex, $\lambda=-1$. Parameters are $n_B/n_F=0.1 $, $k_F a_{BF}=-5$, $k_F a_{FF}=-2/3$.}
\label{NonSelfConsist_Sol}
\end{figure}

When approaching the solution of RG equations in a non-selfconsistent way, i.e without the feedback of the renormalization of the boson-fermion interaction vertex, one encounters a monotonous increase of the  superconducting critical temperature with the bosonic mass. In other words,
if one keeps the $\lambda$-vertex constant in the flow equation for $g$ (see Fig.~ 2(a) in the main text),  one obtains a nonphysical growth of the critical temperature with the mass of bosons in the limit $m_B/m_F\rightarrow\infty$, which is shown in Fig.~\ref{NonSelfConsist_Sol}).

\section{Charge density wave instability}\label{sec:ChargeDensity}
\begin{figure} [h!]
\centering
\includegraphics[width=8cm] {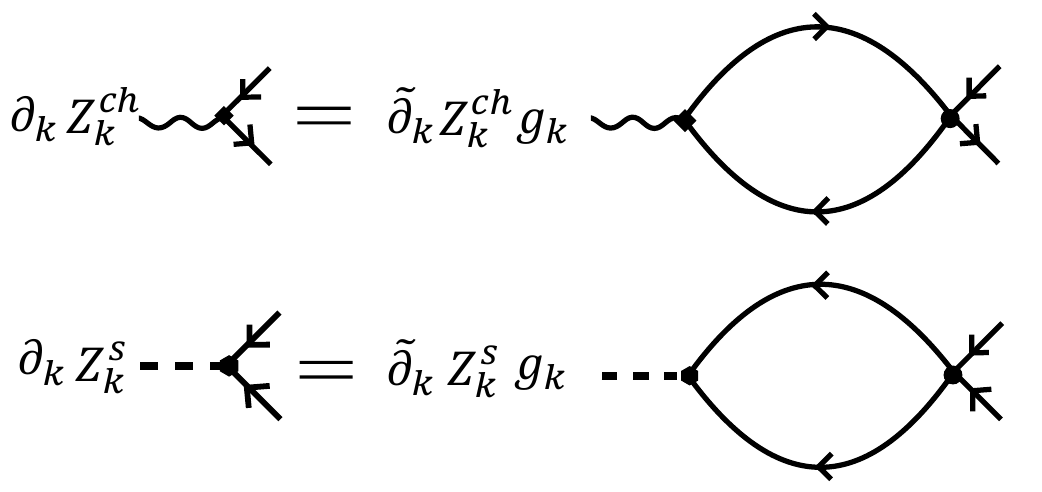}
\caption{Diagrammatic representation of the flow equations of the vertexes $Z_k^{ch}$ and $Z_k^s$ given by Eq.~\eqref{flowCDW}. $g_k$ is the flowing four-fermion point-like coupling constant.}
\label{CDW}
\end{figure}

In order to study if the system is prone to a CDW order instability we compare the  responses of the system to  external fields representing CDW  and superconducting order parameters. To this end we add the source terms 
\begin{equation}
\begin{gathered}
    \Gamma_{source}^{CDW}=\int_{K,Q} Z^{ch}_k h(Q)\left\{\psi^*_{\uparrow}(K+Q)\psi_{\uparrow}(K)+\psi^*_{\downarrow}(K+Q)\psi_{\downarrow}(K)\right\}\:,\\
    \Gamma_{source}^{s}=\int_K Z^s_k J\left[\psi^*_{\downarrow}(K)\psi^*_{\uparrow}(-K) - \psi^*_{\uparrow}(K)\psi^*_{\downarrow}(-K)\right]+\text{h.c.}
\end{gathered}
\end{equation}
to our truncation given by Eq.~(4) of the main text. Comparing the rates at which the vertexes $Z^{ch}_k$ and $Z^{s}_k$ approach infinity, shows which order parameter tends to dominate. The flow equations for $Z^{s}_k$ and $Z^{ch}_k$, which are dependent on the difference of external momenta, e.g. the charge density wave ordering vector, are given by (for a related discussion see, e.g., Ref.~\cite{Desoppi2024})
\begin{equation} \label{flowCDW}
\begin{gathered}
    \frac{\partial Z_k^{ch}}{\partial k}=Z_k^{ch}g_kI_2\:,\\
    \frac{\partial Z_k^{ch}}{\partial k}=Z_k^{ch}g_kI_{BCS}
\end{gathered}
\end{equation}
with the initial conditions $Z_k^{ch}(0)=1$ and $Z_k^{s}(0)=1$. Their diagrammatic representation is shown in Fig.~\ref{CDW}.
These equations are solved together with Eqs.~(5,6) of the main text.

As $I_{BCS}$ is bigger than $I_2$, the vertex corresponding to coupling to the superconducting order parameter approaches infinity faster. Thus  superconductivity will always dominate over the charge density wave state within our model.

\newpage

%